\documentclass[12pt]{JHEP}
\usepackage{amsfonts} 
\usepackage{amsbsy} 
\usepackage{amssymb} 

\newcommand{\C}[1]{{\mathcal #1}}

\newcommand{\beq}{\begin{equation}}
\newcommand{\eeq}{\end{equation}}
\newcommand{\bea}{\begin{eqnarray}}
\newcommand{\eea}{\end{eqnarray}}

\newcommand{\Tr}{{\hbox{Tr}\,}}
\newcommand{\comm}[2]{\left[#1,#2\right]}
\newcommand{\absval}[1]{{\left\vert#1\right\vert}}

\newcommand{\thetafn}[1]{{\,\theta \, \!#1}}
\newcommand{\deltafn}[1]{{\,\delta \, \!#1}}
\newcommand{\expect}[1]{{<#1>}}

\newcommand{\half}{{1\over 2}}

\newcommand{\nn}{\nonumber}

\def\Pf{\C P}
\def\const{ \hbox{\it{const}}}

\title{Convergent Yang-Mills  Matrix Theories}

\author{Peter Austing\\ Department of Physics, University of Oxford \\
Theoretical Physics,\\
1 Keble Road,\\
 Oxford OX1 3NP, UK\\
E-mail: \email{p.austing@physics.ox.ac.uk}}
\author{John F. Wheater\\ Department of Physics, University of Oxford \\
Theoretical Physics,\\
1 Keble Road,\\
 Oxford OX1 3NP, UK\\
E-mail: \email{j.wheater@physics.ox.ac.uk}}

\preprint{hep-th/0103159}

\abstract{We consider the partition function and correlation
functions in the bosonic and supersymmetric Yang-Mills matrix models
with compact semi-simple gauge group. In the supersymmetric case, we
show that the partition function converges when $D=4,6$ and $10$, and that
correlation functions of degree $k< k_c=2(D-3)$ are convergent
independently of the group. In the
bosonic case we show that the partition function is convergent when $D
\geq D_c$, and that correlation functions of degree $k < k_c$ are
convergent, and calculate $D_c$ and $k_c$ for each group, thus extending
our previous results for $SU(N)$.  As a special case these results
establish that the partition function and a set of correlation functions
 in the IKKT IIB string matrix
model are convergent.}


\begin{document}

\section{Introduction}

The quantum mechanics obtained by
dimensional reduction of Yang-Mills field theories to one dimension
was first studied in the 1980s \cite{Claudson:1985th,Savvidy:1985gi} and
describes the sector of the original field
theory in which the fields are independent of the spatial
coordinates. A little later
 it was discovered that the supersymmetric theory (SSYM)
in ten dimensions
with gauge group $SU(N)$ could be regarded in the large $N$
limit as a light-cone regularization of the super-membrane \cite{deWit:1988ig}
but it was soon
realised that  this theory has a continuous spectrum
starting from zero energy \cite{deWit:1989ct}. 
This is in contrast to the case of strings where the spectrum
is discrete, leading to  a tower of states some of which are massless
and therefore candidates for massless and low mass particles in the real
world and the rest of which have masses proportional to the string scale
which is somewhere up in the region of the Planck mass. Thus it appeared
that the absence of a gap rendered the supermembrane theory
 useless as a phenomenological description
of nature and the subject went quiet for many years.

In the ensuing period many remarkable discoveries were made in string
theory. All the known super-string theories  have been found to be
 related
to each other by various duality transformations and to 11-dimensional
supergravity by compactification (see \cite{Green:1999qt,Greene:2000uj,Lechtenfeld:2000vp} for recent reviews and further references), and the
role of D-branes as solitons in string theories opened another window on
non-perturbative string physics  \cite{Polchinski:1995mt}.
  The existence of such  relationships between these
various string
 theories clearly implies that they are all different perturbative
limits of an over-arching theory, christened M-theory. We know M-theory
must be there but we do not know how to write it down.  In particular the 11-dimensional
supergravity is a classical field theory for which a consistent quantization
is not known; however it is contained in the supermembrane theory and this
leads to the BFSS conjecture \cite{Banks:1997vh} that the regularized supermembrane theory
provided by the large $N$ limit of
SSYM quantum mechanics, now christened M(atrix) Theory,
 does indeed represent M-theory in the light-cone gauge. Simultaneously
IKKT  proposed that the dimensional reduction of $D=10$ $SU(N)$ SSYM to zero dimensions
(which is of course the reduction of the quantum mechanics by its remaining
dimension) described in the large $N$ limit the type IIB superstring \cite{Ishibashi:1997xs}.  
For reviews of these subjects see for example \cite{Taylor:2001vb,Aoki:1998bq}.
These developments have
 generated renewed interest in Yang-Mills quantum mechanics
\cite{Sethi:1997pa,Yi:1997eg,Porrati:1998ej,Kac:1999av,Hoppe:1999xg,Staudacher:2000gx}. 
The continuous spectrum is no longer a problem but a virtue because this is
a theory of everything and must describe multi-particle states.  However 
the number of zero energy bosonic states in M(atrix) Theory must be one because there is
only supposed to be one graviton super-multiplet in the theory.  It is
conjectured that this is indeed the case for $SU(N)$ \cite{Witten:1996im} and
calculations in the special case of $SU(2)$ \cite{Sethi:1997pa,Yi:1997eg}
agree but, although there are several partial results \cite{Paban:1998ea,Moore:1998et,Green:1998yf,Konechny:1998vc},
 it remains otherwise unproven. The calculation
of the Witten index requires the evaluation of, among other things,
quantities that are partition functions in the IKKT model;
these are often called Yang-Mills matrix integrals.
 Even here
there are many potential difficulties, the most
basic of which is whether the partition function and 
correlation functions exist
and this is the main subject of this paper.

The  Yang-Mills matrix integral partition function, 
which is obtained by dimensionally reducing the Euclidean SSYM
action from $D$ down to zero dimensions,  is given by
\beq \C Z_{D,G} = \int 
\prod_{\mu=1}^D dX_\mu 
 \prod_{\alpha=1}^{\C{N}} d \psi_{\alpha}
\exp \left( \sum_{\mu , \nu}
\Tr \comm{X_\mu}{X_\nu}^2 + \Tr \psi_{\alpha} [ \Gamma_{\alpha
\beta}^{\mu} X_{\mu}, \psi_{\beta}] \right) \label{P1}
\eeq
where we adopt the summation convention for repeated indices.
The traceless hermitian matrix fields $X_\mu$ and $\psi_\alpha$ (respectively bosonic and
fermionic) are in the Lie algebra $\C G$ of the (compact semi-simple)
 gauge group $G$ and can be written
\beq 
X_\mu=\sum_{a=1}^g X_\mu^a t^a \; , \;\;\;\; \psi_{\alpha} = \sum_{a=1}^g
\psi_{\alpha}^a t^a 
\label{P2}
\eeq
where $\{t^a, a=1,\ldots ,g\}$ are the generators in the fundamental
representation. The $\Gamma^\mu_{\alpha \beta}$ are ordinary gamma
matrices for $D$ Euclidean dimensions. The model possesses a gauge symmetry
\beq X_\mu\to U^\dagger X_\mu U,\quad\psi_\alpha\to U^\dagger \psi_\alpha U,\qquad U\in G. \label{gaugetr}\eeq
and an $SO(D)$ symmetry inherited from the original $D$-dimensional
Euclidean symmetry of the SSYM. 
 Although the  motivation discussed above leads to a study
of the $D=10$, $SU(N)$ supersymmetric integral, it is useful and illuminating to
study several different versions  of the model. Firstly by suppressing
the fermions we get the bosonic integrals which we will denote by $\C N=0$
(ie there are no super-charges)  \cite{Krauth:1998yu}. 
Secondly the supersymmetric integrals can be written
 for $D=3$, 4, 6, and 10, having
$\C N=2(D-2)$ super-charges.
In principle one can integrate out the fermions to obtain
\beq
\label{1.3}
\C Z_{D,G} =\int 
\prod_{\mu=1}^D dX_\mu \,
\Pf_{D,G} (X_\mu )
\exp \left( \sum_{\mu ,\nu}
\Tr \comm{X_\mu}{X_\nu}^2 \right)
\eeq
where the Pfaffian $\Pf_{D,G}$ is a homogeneous polynomial of degree
$\half \C N g$. 
We will also consider simple   correlation functions of the form
\beq\label{correl}
<\C C_k (X_\sigma )> \;\; = \int 
\prod_{\mu=1}^D dX_\mu \, \C C_k (X_\sigma ) \,
\Pf_{D,G} (X_\mu )
\exp \left( \sum_{\mu ,\nu}
\Tr \comm{X_\mu}{X_\nu}^2 \right)
\eeq
where $\C C_k$ is a polynomial of
degree $k$.

Only when the gauge group is $SU(2)$ is it known how to evaluate all these
integrals in closed analytic form 
\cite{Savvidy:1985gi,Smilga:1986jg,Smilga:1986nt,Sethi:1997pa,Yi:1997eg,Krauth:1998yu,Suyama:1998ig}. We recently established analytically
the convergence criteria for the bosonic integrals in the case of $SU(N)$ \cite{Austing:2001bd}
and in this paper we will extend these results to all other compact
semi-simple gauge groups
and to the supersymmetric integrals. However much has already been learned about the properties
of these integrals by a variety of techniques. The authors of \cite{Moore:1998et} used the supersymmetry to deform the Yang-Mills partition function into a cohomological theory
in which the integrals can be done.  From the point of view of the defining formula
\ref{1.3}  this involves among other things a change of variables and an analytic 
continuation that implicitly depends upon the original integral being convergent.
The method appears to work for the partition function when $D>3$ and gives results
in agreement with numerical calculations at small $N$ \cite{Krauth:1998xh,Krauth:1998yu,Staudacher:2000gx}. 
Unfortunately it 
seems that this method
cannot be used to calculate the many correlation functions that are
 of interest in the original YM model. 
There is no small parameter expansion in these models (the original
gauge coupling $g$ can
be scaled out) but the one-loop effective action can be calculated
as can the $\frac{1}{D}$ expansion 
\cite{Aoki:1998vn,Hotta:1998en,Oda:2000im}; care must be exercised in the 
gaussian approximation because it
 requires a cut-off to be 
introduced even when the full theory is actually convergent.

It was realised in \cite{Krauth:1998xh} that these integrals are amenable to numerical 
calculation for small gauge groups where the Pfaffian can be handled more or less
by brute force; very careful and accurate determination of the integrals for $SU(N)$ with
$N=3$, 4 and 5 \cite{Krauth:1998yu} confirmed the results obtained by deformation (there are a number of subtleties
involved in this comparison arising from the normalization of the measures, in addition
to the problems of actually doing the integrals).
  These authors also examined the bosonic integrals, which were commonly 
believed to be divergent because of the flat directions in the action, and found
that in fact provided $D$ is big enough  they converge too. In later papers these
results were extended to other gauge groups \cite{Krauth:2000bv,Staudacher:2000gx}; as we shall see the conclusions about
convergence contained  therein are entirely  in agreement with the analytic 
results that we explain in this paper.

Another use of numerical simulations is to study correlation functions at much larger
$N$ by Monte Carlo and to try to establish the large $N$ behaviour of the theory.
Simulations for the $SU(N)$ bosonic model up to
$N=768$ have now been reported \cite{Hotta:1998en,Ambjorn:2000dj,Horata:2000ft};
 as can be seen from the analytic bounds \cite{Austing:2001bd}
the convergence properties at large $N$ are very good and a great deal of information
can be obtained.   Intriguingly it even appears that the Wilson loop shows an area
law in a  regime which remains finite as $N$ increases.
It is not quite so easy to study large $N$ for the supersymmetric theories because
of the Pfaffian; brute force evaluation is out of the question. When $D=4$ the Pfaffian
is positive semi-definite and can be expressed as a determinant; this means that it is
possible to deal with the fermions by Monte Carlo using the standard methods of lattice
gauge theory and values of $N$ up to 48 have been studied \cite{Ambjorn:2000bf,Burda:2000mn,Ambjorn:2000dj}.
For $D=6$, 10 the Pfaffian causes real trouble because it is complex and standard
methods do not work. Two ways of coping with this have been tried;
in \cite{Ambjorn:2000dx} the one loop effective action with an ultraviolet cut-off
was simulated using the absolute value of the
Pfaffian while  \cite{Nishimura:2000wf,Nishimura:2000ds} studied
 configurations which
are saddle-points of the phase of the Pfaffian. Although both of these calculations
violate supersymmetry it is interesting that the latter leads to lower dimensional
sub-manifolds dominating the  integral whereas the former does not. The
Yang-Mills quantum mechanics has also been studied in the quenched approximation
\cite{Janik:2000tq} and a supersymmetric random surface model has been simulated
directly \cite{Bialas:2000gf} and compared to the IKKT model.

This paper has two purposes. The first is to show how to extend our
convergence proofs for bosonic
 partition functions and correlators from SU(N)    
to all other compact gauge groups;  in section 2 we show which integrals
converge and in section 3 we conversely show which ones diverge. The 
second purpose is to repeat the exercise for the supersymmetric models
which is done in section 4.
Section 5 is a discussion of our results.

\section{Convergent Bosonic Integrals}

We consider first the integral \ref{P1} without fermions so that $\C N
=0$ and there is no Pfaffian. 
The dangerous regions which might cause \ref{P1} to diverge are where
all the commutators almost vanish but the magnitude of $X_\mu$ goes
to infinity. Hence
we let 
\beq X_\mu=Rx_\mu,\quad \Tr x_\mu x_\mu=1. \label{P4}\eeq
Then
\beq \C Z_{D, G}=\int_0^\infty dR R^{Dg-1} \C X_{D,
G}(R)
\label{P5}\eeq 
where
\beq \C X_{D, G}(R)=\int \prod_{\nu=1}^D dx_\nu \,
\deltafn{\left(1-\Tr x_\mu x_\mu\right)}
\exp\left(-R^4S_{\C G}
\right) \label{P6}\eeq
and
\bea S_{\C G}&=&-\Tr \comm{x_\mu}{x_\nu}\comm{x_\mu}{x_\nu}\nn\\
&=&\sum_{i,j,\mu,\nu}\absval{\comm{x_\mu}{x_\nu}_{i,j}}^2.\label{P6.0}\eea
We note that for any  $R$ the integral $\C X_{D, G}(R)$ is bounded by
a constant  and, if for large $R$
\bea\absval{\C X_{D,
G}(R)}&<&\frac{\const}{R^\alpha},\qquad\hbox{with }\:\alpha> Dg,\label{P6.1}\eea
then the partition function $\C Z_{D, G}$ is finite. Our tactic for proving
convergence of $\C Z_{D, G}$ is  to find a bound of the form
\ref{P6.1} on $\C X_{D, G}(R)$. A sufficient condition for the
correlation function \ref{correl} to converge is obtained by modifying
\ref{P6.1} to require $ \alpha > Dg + k$.

From now on, we are only interested in large $R$, so we shall always
assume $R>1$. Let us split the integration region in \ref{P6} into two
\bea \C R_1({\C G}):&& S_{\C G}
<(R^{-(2-\eta)})^2\nn\\
\C R_2({\C G}):&& S_{\C G}
\ge(R^{-(2-\eta)})^2\label{P7}\eea
where $\eta$ is small but positive.  We see immediately that the contribution
to $\C X_{D, G}(R)$ from $\C R_2({\C G})$ is bounded by $A_1\exp(-R^{2\eta})$ (we will
use the capital letters
$A$, $B$ and $C$ to denote constants throughout this paper)
 and thus automatically satisfies \ref{P6.1}. Thus we can
confine our efforts to the contribution from $\C R_1({\C G})$ in which we replace
the exponential function by unity to get the bound
\beq \absval{\C X_{D, G}(R)}<A_1\exp(-R^{2\eta})+\C I_{D, G}(R) 
 \label{P7.1}\eeq
where 
\beq\C I_{D, G}(R) = \int_{\C R_1({\C G})} \prod_{\nu=1}^D dx_\nu \, \deltafn{\left(1-\Tr x_\mu x_\mu\right)}.
 \label{B8}\eeq 
The condition in \ref{P4} means that at least one of the matrices
$x_\mu$ (say $x_1$) must satisfy
\beq
\label{B9}
\Tr x_1 x_1 \geq D^{-1}.
\eeq
It is convenient to express the Lie algebra $\C G$ using the Cartan-Weyl basis
\beq
\label{B11}
\{ H^i, E^\alpha \}
\eeq
where $i$ runs from $1$ to the rank $l$ and $ \alpha $ denotes
a root.
In this basis
\beq
\comm{H^i}{H^j}=0\, , \;\;\;\; \comm{H^i}{E^\alpha}=\alpha^i E^\alpha
\eeq
and
\beq
\begin{array}{llll}
\comm{E^\alpha}{E^\beta}&=& N_{\alpha
\beta}E^{\alpha+\beta}\;\;\;\;\;\; & \hbox{if } \alpha + \beta \hbox{
is a root}\\
& =&2
\absval{\alpha}^{-2}\, \alpha \cdot H& \hbox{if } \alpha = -\beta \\
&=&0& \hbox{otherwise}
\end{array}
\eeq
Here $E^{-\alpha} = (E^{\alpha})^{\dagger}$, and the normalisation is
chosen  such that 
\beq
\label{innerprod}
\Tr H^i H^j = \delta^{ij} \, , \;\;\; \Tr E^\alpha E^{\beta} =
2  \absval{\alpha}^{-2}\, \delta^{\alpha+ \beta},\qquad \Tr H^iE^\alpha=0.
\eeq
Since the integrand and measure are gauge invariant, we can always make
a gauge transformation \ref{gaugetr}
to move $x_1$ into the Cartan subalgebra
\beq
x_1 = x^i H^i
\eeq
and reduce the integral over $x_1$ to an
integral over its Cartan modes \cite{Krauth:2000bv}
\beq
\prod_{a=1}^{g} dx_1^a \rightarrow \const \left( \prod_{i=1}^l dx^i
\right) \Delta^2_G(x)\eeq
where the Weyl measure
\beq
\label{vandermonde}
\Delta^2_G(x) = \prod_{\alpha > 0} ( x \cdot \alpha
)^2
\eeq
is the generalisation from $SU(N)$ of the Vandermonde determinant factors. We
expand the remaining $x_\nu$
\beq
x_\nu = x_\nu^i H^i + x_\nu^\alpha E^\alpha \, \;\;\;\; \nu=2, \cdots
, D
\eeq
with $x_\nu^{-\alpha} = (x_\nu^\alpha)^*$.

In the region $\C R_1({\C G})$, we certainly have $-  \Tr
\comm{x_1}{x_\nu}^2 < R^{-2(2-\eta )}$ for $\nu=2 , \cdots ,D$, and
writing this in terms of the basis \ref{B11} gives 
\beq
\label{B15}
4\sum_{\alpha >0} \frac{(x \cdot \alpha)^2}{\absval{\alpha}^2} \absval{ x_\nu^\alpha}^2 <
R^{-2(2-\eta )},
\eeq
where the sum is over positive roots.
This is the key result because, 
whenever $(x \cdot \alpha )^2$ is bigger than a constant, it gives us
a bound of order $R^{-(2-\eta)}$ on $x_\nu^\alpha$ and so allows us to bound the integral \ref{B8}.

There is only a finite number of ways of choosing the positive
roots and
it is convenient 
to define them in the following manner.
 We partition all the roots according to whether
 $x\cdot\alpha$ is i) positive, in which case we call $\alpha$ a positive root,
ii)  negative, in which case we call $\alpha$ a negative root, iii) zero, which
we call the orthogonal subspace. 
Finally we partition the  roots in the orthogonal subspace 
into positive or negative by the standard  Cartan construction.
Thus we see that if $\alpha$ is a positive root we  have by construction
$x\cdot\alpha\ge 0$. As usual there is a set of $l$ simple positive
roots $\{ s_i \}$ which span the $l$-dimensional root space
and any positive root can be be written 
\beq\alpha =\sum_{i=1}^l
n^\alpha_i s_i, \label{rootdec}\eeq
 with the $\{ n^\alpha_i \}$ non-negative integers. 

The integration region of $x$ is split into a finite number of
sub-regions; one for each choice of the positive roots. In each
sub-region, the properties discussed in the previous paragraph hold. 
However, since all possible sets of simple positive roots are related
by Weyl reflections, each of these sub-regions is equivalent as far
as our integrals are concerned.
Now
define a number $c$ 
\beq
c = \min_{ \{a^2=1\}}  \, \max_{i} \, \absval{a \cdot
s_i}
\eeq
which must be positive ($c$ can be related to the quadratic form
matrix but we do not need an explicit expression).
Then the
condition \ref{B9} tells us that at least one of the simple roots,
$s_1$ say, satisfies $x \cdot s_1 \geq c D^{-\half}$. In addition, any
positive root $\alpha$ which contains the simple root $s_1$ satisfies
$x \cdot \alpha \geq c D^{-\half}$ on account of \ref{rootdec}.

We now split up the Lie algebra $\C G$ as follows.
Define $\C G' = Lie(G')$ to be the regularly embedded subalgebra of $\C
G$ obtained by omitting the simple root $s_1$. Then $\hbox{rank}(\C
G') = \hbox{rank}(\C G)-1$ and we write $J$ for the combination of 
the generators $H^i$ which commutes with
$\C G'$.  The remaining generators are $\{ F^\beta \}$
where $\beta$ is any root which contains $s_1$.  As a simple
consequence of the root 
structure and
construction of $\C G'$, we note
\beq
\label{bases}
\begin{array}{lll}
\comm{J}{\C G'}&=&0\\
\comm{F^\beta}{\C G'}& \subset& \{ F^\gamma \}\\
\comm{J}{F^\beta}& \subset&\{ F^\gamma \}\\
\comm{\C G'}{\C G'} &\subset&\C G',
\end{array}
\eeq
and then decompose $x_\mu$ into
\beq
\label{xdecomp}
x_\mu = y_\mu + \rho_\mu J + \omega_\mu^\beta F^\beta,
\eeq
with $y_\mu \in \C G'$;
the condition \ref{B15} then gives us a bound
on the $\omega_\mu$,
\beq
\label{B16}
\absval{\omega_\nu^\beta} < c^{-1} D^{\half}
R^{-(2-\eta)} \, , \;\;\;\nu = 2, \cdots ,D.
\eeq
We must further split up the
integration region according to the relevant choice of $G'$, and then
we 
can use \ref{B16} to bound the integral \ref{B8} in each of 
these regions. The region giving the least inverse power of $R$ will
then give a bound on $\C I_{D, G}$. Using the decomposition
\ref{xdecomp}, the commutation rules \ref{bases}, and the inner products
\ref{innerprod}, we find that the action takes the form
\bea
\label{expaction}
S_{\C G}(x_\mu) &=& S_{\C G'}(y_\mu) + 2\Tr\comm{y_\mu}{y_\nu}\comm{F^\beta}{F^\gamma}
\omega^\beta_\mu \omega^\gamma_\nu\nn\\
&&+\Tr(\omega^\beta_\nu\comm{y_\mu}{F^\beta}-\omega^\beta_\mu\comm{y_\nu}{F^\beta}
+(\rho_\mu\omega_\nu^\beta-\rho_\nu\omega_\mu^\beta)\comm{J}{F^\beta}
+\omega^\beta_\mu \omega^\gamma_\nu \comm{F^\beta}{F^\gamma})^2.\nn\\
&=& S_{\C G'}(y_\mu)+ \C O(R^{-2(2-\eta)})
\eea
where we have used \ref{B16} and the fact that the elements of $y_\mu$ and
 $\rho_\mu$ are 
bounded by a constant. Thus
we find that (up to a trivial scaling constant)
\beq
\label{B18}
x_\mu \in \C R_1(\C G) \Rightarrow y_\mu \in \C R_1(\C G').
\eeq
The final ingredient is to note that the Weyl measure
\ref{vandermonde} for 
$ G$ can be bounded by that for $ G'$:
\beq\label{B20}
\Delta^2_G(x) < \const \, \Delta^2_{G'}(y)
\eeq
Then, integrating out the $\omega$ and $\rho$ degrees of freedom,
and using the bounds \ref{B16}, \ref{B18} and \ref{B20} gives
 (more details of these manipulations are given in our previous
paper \cite{Austing:2001bd})
\beq\label{B21}
\C I_{D, G}(R) < B_1 R^{-(2-\eta)(D-1)(g-g' -1)} \C F_{D,G'}(R)
\eeq
where
\beq
\C F_{D,G'}(R) =
\int_{\C
R_1(G')} \prod_{\nu=1}^D dy_\nu \, 
\thetafn{\left(1-\Tr y_\mu y_\mu\right)},
\eeq
and we have absorbed the $G'$ Weyl measure thus  restoring the integral
to $G'$ gauge invariant form.
Using the identity
$
\thetafn{ (1 - \Tr y_\mu y_\mu) }= \int_0^1 dt
\deltafn{(t - \Tr y_\mu y_\mu) }
$
and then rescaling $t =
[u/R]^{2-\eta}$ and $y_\mu = \tilde{y}_\mu [u/R]^{1-\eta /2}$ gives
\beq\label{B22}
\C F_{D, G'}(R) = (2-\eta ) R^{-(1-\eta /2)Dg'}
\int_0^R du  \,u^{(1-\eta /2)Dg'-1} \C I_{D,G'}(u).
\eeq
We shall proceed by induction. Our aim is to show that
\beq
\int_0^\infty dR \, R^{Dg-1} \C I_{D,G}(R) < const.\label{bounded}
\eeq 
 If this is true for $G'$, then \ref{B22} tells us
\beq\label{BF}
\C F_{D, G'}(R) < B_2 R^{-(1-\eta /2)Dg'}
\eeq
so that by \ref{B21}
\beq\label{B24}
\C I_{D,G}(R) < B_3 R^{-(1-\eta /2)[2(D-1)(g-g'-1) + Dg']}
\eeq
and we can then decide on the truth of \ref{bounded} for $G$.
Our task then is to find the regularly embedded subalgebras $\C G'$ of $\C G$
and choose the one which leads to the least inverse power in
\ref{B24}. In doing this we will need the result that 
if the regularly embedded subalgebra $\C G'$ is a direct
sum of two (mutually commuting) subalgebras $\C G' = \C G'_1 \oplus \C
G'_2$ then 
\beq
\C F_{D,G'}(R) < \C F_{D,G'_1}(R) \, \C F_{D,G'_2}(R)
\eeq
since $ \thetafn{(1-\Tr_{\C G'} y_\mu y_\mu )} \leq \thetafn{(1-\Tr_{\C G'_1}
y_\mu y_\mu )} \thetafn{(1-\Tr_{\C G'_2} y_\mu y_\mu )}$ and
$S_{\C G'}=S_{\C G'_1}+S_{\C G'_2}$.

We now proceed to consider each group in turn:

\begin{description}
\item[SU(r+1):]
The case of $SU(r+1)$ is dealt with in our previous paper
\cite{Austing:2001bd}. We include a review here for completeness.
The Dynkin diagram for $su(r+1)$ is
\beq
\unitlength=1mm
\linethickness{0.4pt}
\begin{picture}(64,6)
\put(2,2){\circle{4}}
\put(17,2){\circle{4}}
\put(32,2){\circle{4}}
\put(47,2){\circle{4}}
\put(62,2){\circle{4}}
\put(4,2){\line(1,0){11}}
\put(19,2){\line(1,0){11}}
\put(37,2){\circle*{0.53}}
\put(39.5,2){\circle*{0.53}}
\put(42,2){\circle*{0.53}}
\put(49,2){\line(1,0){11}}
\end{picture}
\eeq
where there are $r$ nodes. To find
the regularly embedded subalgebras $\C G'$
we simply remove one of the nodes, and discover
\beq
su(r+1) \rightarrow \C G' = su(m) \oplus su(r+1-m) \, , \;\;\; 1 \leq
m \leq r. 
\eeq
where we define $su(1) = 0$.
The dimension of $su(m)$ is $m^2-1$, so that $g' = m^2 +(r+1-m)^2 -2$.

The Lie algebra $su(2)$ has no regularly embedded subalgebra, so
$g'=0$ and
\beq
\C I_{D,SU(2)} < B_3 R^{-(1-\eta /2)4(D-1)}.
\eeq
Then $\C Z_{D,SU(2)}$ is finite for $D \geq 5$. Substituting into
\ref{B22}, we see
\beq
\label{B29}
\C F_{D,SU(2)} < B_4 R^{(1-\eta /2)3D} R^{(1-\eta /2) \delta_{D,3}}
(\log R)^{\delta_{D,4}}.
\eeq
In dimensions $3$ and $4$, the result is at variance with
\ref{BF}. However, modification by a $\log R$ factor will not affect
any of our conclusions, so it is only for $D=3$ that we must be
careful to use the modified formula.

The Lie algebra $su(3)$ has $su(2)$ as its only regularly embedded
subalgebra. Then substituting \ref{B29} into \ref{B21}, we discover
$\C Z_{D,SU(3)}$ converges for $D \geq 4$. In this case,  the general
formula \ref{BF} is modified only in the case $D=3$, 
and only by a factor of $\log R$ which will not affect our results.

For $SU(r+1)$ with $r \geq 3$, it is a simple exercise to discover
which of these possible $\C G'$ 
gives the least inverse power of $R$ behaviour in \ref{B24}. The only
point to remember is that we must include an extra $R^{1-\eta/2}$
factor in the case of $\C G' = su(2) \oplus su(r-1)$ when $D=3$, to
allow for the anomalous behaviour of $\C F_{3,SU(2)}$.
We discover that we must always take $\C G' = su(r)$, giving $g'=
r^2-1$. Substituting back into \ref{B24}, we find that $\C
Z_{D,SU(r+1)}$ is convergent for $D \geq 3$ when $r \geq 3$.
The correlation function \ref{correl} converges when $k<k_c$ with
\beq
k_c = 2rD -D -4r-\delta_{D,3}\delta_{r,2}\;,\qquad r\ge 2, D\ge 3.
\eeq

\item[SO(2r+1), $\bf{r \geq 2}$:]
The Dynkin diagram for $so(2r+1)$ is
\beq
\unitlength=1mm
\linethickness{0.4pt}
\begin{picture}(64,6)
\put(2,2){\circle*{4}}
\put(17,2){\circle{4}}
\put(32,2){\circle{4}}
\put(47,2){\circle{4}}
\put(62,2){\circle{4}}
\put(3.732,3){\line(1,0){11.536}}
\put(3.732,1){\line(1,0){11.536}}
\put(19,2){\line(1,0){11}}
\put(37,2){\circle*{0.53}}
\put(39.5,2){\circle*{0.53}}
\put(42,2){\circle*{0.53}}
\put(49,2){\line(1,0){11}}
\end{picture}
\eeq
where there are $r$ nodes, and the dimension is $g=2r^2+r$. By
removing one  node, we see 
that the possible $\C G'$ are $so(2m+1) \oplus su(r-m)$ with $0 \leq m
\leq r-1$. We discover the most important contribution is always from
$\C G' = so(2r-1)$, and that $\C Z_{D,SO(2r+1)}$ always converges
for $r \geq 2$ and $D \geq 3$. The critical degree $k_c$ for
correlation functions is
\beq
\begin{array}{lllll}
k_c &=& 2&\;\;\;\;&r=2 \, , D=3 \\
k_c &=& 4rD-8r-3D+4&\;\;\;\;&\hbox{otherwise}.
\end{array}
\eeq
The exception when $r=2$ and $D=3$ occurs because of the anomalous
behaviour of $\C F_{3,SU(2)}$.

\item[Sp(2r), $\bf{r \geq 2}$:]
The Dynkin diagram for $sp(2r)$ is
\beq
\unitlength=1mm
\linethickness{0.4pt}
\begin{picture}(64,6)
\put(2,2){\circle{4}}
\put(17,2){\circle*{4}}
\put(32,2){\circle*{4}}
\put(47,2){\circle*{4}}
\put(62,2){\circle*{4}}
\put(3.732,3){\line(1,0){11.536}}
\put(3.732,1){\line(1,0){11.536}}
\put(19,2){\line(1,0){11}}
\put(37,2){\circle*{0.53}}
\put(39.5,2){\circle*{0.53}}
\put(42,2){\circle*{0.53}}
\put(49,2){\line(1,0){11}}
\end{picture}
\eeq
where there are $r$ nodes, and the dimension is $g=2r^2+r$. The
possible $\C G'$ are $sp(2m) \oplus su(r-m)$ with $0 \leq m
\leq r-1$, and the dominant contribution is from $sp(2r-2)$. The
partition function $\C Z_{D,Sp(2r)}$ converges for all $r \geq 2$ and
$D \geq 3$ and the critical correlation function is given by
\beq 
\begin{array}{lllll}
k_c &=& 2&\;\;\;\;&r=2 \, , D=3 \\
k_c &=& 4rD-8r-3D+4&\;\;\;\;&\hbox{otherwise}.
\end{array}
\eeq

\item[SO(2r), $\bf{r \geq 4}$:]
The Dynkin diagram for $so(2r)$ is
\beq
\unitlength=1mm
\linethickness{0.4pt}
\begin{picture}(64,18)
\put(2,2){\circle{4}}
\put(17,2){\circle{4}}
\put(32,2){\circle{4}}
\put(47,2){\circle{4}}
\put(62,2){\circle{4}}
\put(47,14){\circle{4}}
\put(4,2){\line(1,0){11}}
\put(22,2){\circle*{0.53}}
\put(24.5,2){\circle*{0.53}}
\put(27,2){\circle*{0.53}}
\put(34,2){\line(1,0){11}}
\put(49,2){\line(1,0){11}}
\put(47,4){\line(0,1){8}}
\end{picture}
\eeq
where there are $r$ nodes, and the dimension is $g=2r^2-r$. The
possible $\C G'$ are $so(2m) \oplus su(r-m)$ for $4 \leq m \leq r-1$,
$su(4) \oplus su(r-3)$, $su(r-2) \oplus su(2) \oplus su(2)$ and
$su(r)$. The dominant contribution always comes from $so(2r-2)$, and we
discover that $\C Z_{D,SO(2r)}$ always converges for $D \geq 3$ and $r
\geq 4$. The critical correlation function is given by
\beq
k_c = 4rD -5D -8r +8.
\eeq

\item[$\bf{G_2}$:]
The Dynkin diagram is
\beq
\unitlength=1mm
\linethickness{0.4pt}
\begin{picture}(64,6)
\put(2,2){\circle{4}}
\put(17,2){\circle*{4}}
\put(4,2){\line(1,0){11}}
\put(3.323,0.5){\line(1,0){12.354}}
\put(3.323,3.5){\line(1,0){12.354}}
\end{picture}
\eeq
and the dimension is $14$. The only regularly embedded subalgebra is
$su(2)$, and we discover $\C Z_{D,G_2}$ converges for $D \geq 3$ with
\beq
k_c = 9D - 20 - \delta_{D,3}\;.
\eeq

\item[$\bf{F_4}$:]
The Dynkin diagram is
\beq
\unitlength=1mm
\linethickness{0.4pt}
\begin{picture}(64,6)
\put(2,2){\circle*{4}}
\put(17,2){\circle*{4}}
\put(32,2){\circle{4}}
\put(47,2){\circle{4}}
\put(4,2){\line(1,0){11}}
\put(18.732,3){\line(1,0){11.536}}
\put(18.732,1){\line(1,0){11.536}}
\put(34,2){\line(1,0){11}}
\end{picture}
\eeq
and the dimension $g=52$. The dominant contributions come equally from
$\C G' = so(7)$ and $\C G' = sp(6)$, each having $g'=21$. Then $\C
Z_{D,F_4}$ converges for $D 
\geq 3$ and
\beq
k_c = 29D-60.
\eeq

\item[$\bf{E_6}$:]
The Dynkin diagram is
\beq
\unitlength=1mm
\linethickness{0.4pt}
\begin{picture}(64,18)
\put(2,2){\circle{4}}
\put(17,2){\circle{4}}
\put(32,2){\circle{4}}
\put(47,2){\circle{4}}
\put(62,2){\circle{4}}
\put(32,14){\circle{4}}
\put(4,2){\line(1,0){11}}
\put(19,2){\line(1,0){11}}
\put(34,2){\line(1,0){11}}
\put(49,2){\line(1,0){11}}
\put(32,4){\line(0,1){8}}
\end{picture}
\eeq
and the dimension $g=78$. The dominant contribution comes from $\C G'
= so(10)$ having $g'=45$. Then $\C Z_{D,E_6}$ converges for $D \geq 3$
and
\beq
k_c = 31D-64.
\eeq

\item[$\bf{E_7}$:]
The Dynkin diagram is
\beq
\unitlength=1mm
\linethickness{0.4pt}
\begin{picture}(79,18)
\put(2,2){\circle{4}}
\put(17,2){\circle{4}}
\put(32,2){\circle{4}}
\put(47,2){\circle{4}}
\put(62,2){\circle{4}}
\put(77,2){\circle{4}}
\put(32,14){\circle{4}}
\put(4,2){\line(1,0){11}}
\put(19,2){\line(1,0){11}}
\put(34,2){\line(1,0){11}}
\put(49,2){\line(1,0){11}}
\put(64,2){\line(1,0){11}}
\put(32,4){\line(0,1){8}}
\end{picture}
\eeq
and the dimension $g=133$. The dominant contribution comes from $\C G'
=e_7$ with $g'=78$. Then $\C Z_{D,E_7}$ converges for $D \geq 3$ and
\beq
k_c = 53D-108.
\eeq

\item[$\bf{E_8}$:]
The Dynkin diagram is
\beq
\unitlength=1mm
\linethickness{0.4pt}
\begin{picture}(94,18)
\put(2,2){\circle{4}}
\put(17,2){\circle{4}}
\put(32,2){\circle{4}}
\put(47,2){\circle{4}}
\put(62,2){\circle{4}}
\put(77,2){\circle{4}}
\put(92,2){\circle{4}}
\put(32,14){\circle{4}}
\put(4,2){\line(1,0){11}}
\put(19,2){\line(1,0){11}}
\put(34,2){\line(1,0){11}}
\put(49,2){\line(1,0){11}}
\put(64,2){\line(1,0){11}}
\put(79,2){\line(1,0){11}}
\put(32,4){\line(0,1){8}}
\end{picture}
\eeq
with dimension $g=248$. The dominant contribution comes from $\C G' =
e_7$ with $g'=133$. Then $\C Z_{D,E_8}$ converges for $D \geq 3$ and
\beq
k_c = 113D-228.
\eeq
\end{description}

\section{Divergent Bosonic Integrals}
The
partition functions
 trivially diverge when $D=2$ since,
taking $X_1$ to be in the Cartan subalgebra, the integrand is
independent of the Cartan subalgebra degrees of freedom of $X_2$.
In \cite{Austing:2001bd} we showed that the $SU(N)$ partition
function diverges whenever the convergence conditions are not met. 
We will now show that the correlation function $\expect{(\Tr X_\mu X_\mu)^{k/2}}$
always diverges when $k\ge k_c$ so that $k_c$ is indeed critical.

 We
have immediately that 
\beq \expect{(\Tr X_\mu X_\mu)^{k/2}}=\int_0^\infty dR R^{Dg+k-1}\C X_{D,G}(R).
\eeq
 This time, consider the region
\bea \C R:&& S
<R^{-4}\eea
Then $\exp (-R^4 S) > \exp(-1)$ and so
\beq
\label{D2}
\C X_{D, G}(R) >C_1 \C I_{D,G}
\eeq
where now
\beq
\C I_{D,G}=\int_{\C R} \prod_{\nu=1}^D dx_\nu \,
\deltafn{\left(1-\Tr x_\mu x_\mu\right)}
\eeq
and, moving $x_1$ into the Cartan subalgebra,
\beq
\label{D3}
\C X_{D, G}(R) > C_2 \int_{\C R} \prod_{i=1}^l dx_1^i
\Delta^2_G(x_1^i)  \prod_{\nu=2}^D dx_\nu \, 
\deltafn{\left(1-\Tr x_\mu x_\mu\right)}.
\eeq
Now pick a regularly embedded sub-algebra $\C G'$ of $\C G$ (with rank
$1$ less than $\C G$).
As before, write $x =y+
\rho J  + \omega^\beta F^\beta$ with $y \in \C G'$, and define a new region $\C R'_\epsilon$ by
\beq
\begin{array}{lll} 
\C R'_\epsilon: \;\;\;& \absval{\omega_\nu^\beta} < \epsilon
R^{-2}& \nu = 2, \cdots ,D \\
& S_{G'}(y_\mu)  <
\epsilon R^{-4}.
\end{array} 
\eeq
Then by
taking $\epsilon$ small enough, we see from \ref{expaction} that 
\beq
R'_\epsilon \subset \C R
\eeq
and therefore
\beq
\label{D4}
\C X_{D, G}(R) > C_2 \int_{\C R'_\epsilon} \prod_{i=1}^l dx_1^i
\Delta^2_G(x_1^i)  \prod_{\nu=2}^D dx_\nu \, 
\deltafn{\left(1-\Tr x_\mu x_\mu\right)}.
\eeq
Now perform this integral over the $\omega$ and $\rho$ to discover
\beq
\label{D8}
\C X_{D, G}(R) > C_3 R^{-2(D-1)(g-g'-1)} \C F_{D,G'}(R)
\eeq
where in this case
\bea
\C F_{D,G}(R) &=& \int_{\C R}
\prod_{\nu=1}^D dx_\nu \,
\thetafn{\left(1-\Tr x_\mu x_\mu\right)} \left(1-\Tr x_\mu x_\mu\right)^{(D-1)/2} \\
&=& C_4 R^{-Dg}  \int_0^R du u^{Dg-1} \left(1-\frac{u}{R}\right)^{(D-1)/2}\C I_{D,G}(u). 
\eea
If $\C G'$ does not contain $su(2)$ as an invariant subalgebra, then
\beq\int_0^R du u^{Dg-1}\left(1-\frac{u}{R}\right)^{(D-1)/2} \C I_{D,G}(u) > C_5=\const \;\;\; (\hbox{for } R> 1).
\eeq 
In the case of $SU(2)$, we can repeat the argument leading to \ref{D8}
(taking $G'=1$) to find $\C I_{D,SU(2)} > C_5 u^{-4(D-1)}$ for large
$u$ so that
\beq
\begin{array}{llll}
\int_0^R du u^{Dg-1} \left(1-\frac{u}{R}\right)^{(D-1)/2}\C I_{D,G}(u) & > & C_6 & D \geq 5\\
&> \;\;\;& C_7 \log R & D=4 \\
&>& C_8 R & D=3
\end{array}
\eeq
Then 
\beq
\C X_{D, G}(R) > C_9 R^{-2(D-1)(g-g'-1)} R^{-Dg'}
R^\delta,
\eeq
where $\delta=1$ if $D=3$ and $G'=SU(2)$ and zero otherwise,
which is essentially the converse of \ref{B24}.
Finally the usual power counting argument leads to the result
(note that this time we do 
not need to go through all the sub-algebras but
 just the one which gives the most divergent 
behaviour for the partition function).

\section{Convergent Supersymmetric Integrals}
We proceed as for the bosonic integrals to set
\beq \C Z_{D, G}=\int_0^\infty dR  R^{Dg-1} R^{(D-2)g} \C X_{D,
G}(R)
\eeq 
where now
\beq \C X_{D, G}(R)=\int \prod_{\nu=1}^D dx_\nu \, \C P_{D,G}(x_\sigma)
\deltafn{\left(1-\Tr x_\mu x_\mu\right)}
\exp\left(-R^4S_{\C G}\right). \eeq
As before, it is sufficient to consider the region
\beq
\C R_1(\C G):\;\;\; S_{\C G}< R^{-2(2-\eta)}
\eeq
We shall again argue by induction, and
for the induction step to work, we actually need to prove the result for the
generalised Pfaffian
\beq
\label{modpfaff}
[\Pf^r_{D,G}(x,R)]^{a_1,\cdots ,a_{2r}}_{\alpha_1,\cdots ,\alpha_{2r}}
= R^{-(2-\eta )2r} \int d \psi \exp (\Tr
{\Gamma}^{\mu}_{\alpha \beta} \psi_\alpha [x_\mu ,
\psi_\beta ] ) \psi^{a_1}_{\alpha_1} \cdots \psi^{a_{2r}}_{\alpha_{2r}}
\eeq
which exists for $r=0,\ldots (D-2)g$, and 
is modified from the
usual definition by the inclusion of $2r$ fermionic insertions, each with
an accompanying 
factor of $R^{-(2-\eta )}$.
 If we set 
$r=0$ then the
original Pfaffian $\Pf_{D,G}$ is recovered (and is of course
independent of $R$).

The structure of the ${\Gamma}$s will be irrelevant from now
on; their only relevant property, which we will use repeatedly, is that
$\absval{\Gamma^\mu_{\alpha\beta}}\le 1$. For a more compact notation we shall suppress the
dependence on ${\Gamma}$, and the explicit spinor and group  indices, and write
\beq
\label{simplepfaff}
\Pf^r_{D,G}(x,R)
= R^{-(2-\eta )2r} \int d \psi \exp (\Tr
\psi [x ,
\psi ] ) \psi^1 \cdots \psi^{2r}.
\eeq 
Then defining
\beq\C I^r_{D, G}(R) = \int_{\C R_1(\C G)} \prod_{\nu=1}^D dx_\nu \,
\absval{\Pf^r_{D,G} 
(x,R)} \deltafn{\left(1-\Tr x_\mu x_\mu\right)}\label{Idef}
\eeq 
we have
\beq 
\absval{\C X_{D, G}^r(R)}<A_1\exp(-R^{2\eta})+\C I_{D, G}^r(R).
\eeq 
Proceeding as in the bosonic case, we choose the relevant regularly
embedded subalgebra $\C G'$, expand $x_\mu = y_\mu + \rho_\mu J +
\omega_\mu^\beta F^\beta$, and note
\beq
\absval{\omega_\nu^\beta} < c^{-1} D^{\half}
R^{-(2-\eta)} \, , \;\;\;\nu = 2, \cdots ,D.
\eeq
Further, write
\beq
\psi =\phi + \xi +  \chi
\eeq
with $\phi \in \C G'$, $\xi = \xi J$ and $\chi = \chi^\beta F^\beta$.
Using the relations \ref{bases}, we find
\beq\label{F12}
\begin{array}{lll}
\Tr \psi \comm{x}{\psi}& =& \Tr \phi \comm{y}{\phi} \\
&&+ \Tr \phi
\comm{\omega}{\chi}+
\Tr \chi \comm{\omega}{\phi} \\
&&
+ \Tr \chi \comm{\omega}{\xi}
+ \Tr \xi \comm{\omega}{\chi}\\
&&+ \Tr \chi
\comm{x}{\chi}
\end{array}
\eeq
where $\rho = \rho J$ and $\omega= \omega^\beta F^\beta$.
Inserting this expression into the definition \ref{modpfaff}, 
and expanding part of the exponential,  we get
\bea \C P^r_{D,G}(x,R)&=&\int d\phi d\chi d\xi
\left(\frac{\xi^1\cdots\xi^k}{R^{k(2-\eta)}}
\frac{\phi^1\cdots\phi^m}{R^{m(2-\eta)}}
\frac{\chi^1\cdots\chi^n}{R^{n(2-\eta)}}\right)\nn\\
&&\times\exp\left(\Tr \phi \comm{y}{\phi} 
+\Tr \phi\comm{\omega}{\chi}+\Tr \chi \comm{\omega}{\phi} 
+ \Tr \chi\comm{x}{\chi}\right)\nn\\
&&\times\frac{1}{(2(D-2)-k )!}\left( \Tr \chi \comm{\omega}{\xi}
+ \Tr \xi \comm{\omega}{\chi}\right)^{2(D-2)-k},
\eea
where $k+m+n=2r$.
First we do the integrals over the 
$\C N =2(D-2)$ grassman variables $\xi_\alpha$ each of 
which is paired either with an $\omega$, or with an explicit factor
$R^{-(2-\eta)}$; since $\omega$ is itself bounded we get
\bea \absval{\C P^r_{D,G}(x,R)}&<&\frac{R^{-2(D-2)(2-\eta)}}{(2(D-2)-k)!}
\sum_{P} \Bigg\vert \int d\phi  d\chi
\left(\frac{\phi^1\cdots\phi^m}{R^{m(2-\eta)}}
\frac{\chi^1\cdots\chi^{n+2(D-2)-k}}{R^{n(2-\eta)}}\right)\nn\\
&&\times \exp\left(\Tr \phi \comm{y}{\phi} 
+\Tr \phi\comm{\omega}{\chi}+\Tr \chi \comm{\omega}{\phi} 
+ \Tr \chi\comm{x}{\chi}\right)\Bigg\vert\eea
where $P$ simply indicates all the possible permutations of indices
that can be generated. Next we expand the $\phi\omega\chi$ terms to get
\bea \absval{\C P^r_{D,G}(x,R)}&<&R^{-2(D-2)(2-\eta)}
\sum_{P}\sum_l\frac{2^l}{l! (2(D-2)-k)!}\nn\\
&&\times \absval{\int d\phi
\left(\frac{\phi^1\cdots\phi^{m+l}}{R^{(m+l)(2-\eta)}}
\right)\exp\left(\Tr \phi \comm{y}{\phi}\right) }\nn\\
&&\times\max_x\absval{\int d\chi\left(\frac{\chi^1\cdots
\chi^{n+2(D-2)-k+l}}{R^{n(2-\eta)}}\right)
\exp\left( \Tr \chi\comm{x}{\chi}\right)}.\eea
Finally we integrate out the $\chi$ fermions and use the fact that $x$
is a bounded quantity to obtain
\beq
\absval{ \C P^r_{D,G}(x,R)} < R^{-(2-\eta)2(D-2)} \sum_{r'} C_{r'}
\absval{ \C P^{r'}_{D,G'}(y,R) }\label{Pbound}
\eeq
where the $C_{r'}$ are constants. In the spirit of the notation
\ref{simplepfaff}, we have suppressed sums over the many possible
combinations of
 indices.

Inserting the bound \ref{Pbound} into \ref{Idef} we get
\beq
\C I^r_{D, G}(R) < R^{-(2-\eta)2(D-2)} \sum_{r'} C_{r'} \int_{\C R_1} \prod_{\nu=1}^D dx_\nu \,
\absval{\Pf^{r'}_{D,G'} 
(y,R)} \deltafn{\left(1-\Tr x_\mu x_\mu\right)}
\eeq 
and we can now follow the bosonic procedure and integrate out the
$\omega$ and $\rho$ degrees of freedom to obtain
\bea \C I^r_{D, G}(R)& <& R^{-(2-\eta)2(D-2)} R^{-(2-\eta)(D-1)(g-g'-1)}\nn \\
&&\times \sum_{r'} C_{r'} \int_{\C R_1(G')} \prod_{\nu=1}^D dy_\nu \, 
\absval{\Pf^{r'}_{D,G'} 
(y,R)} \thetafn{\left(1-\Tr y_\mu y_\mu \right)}.
\eea
As before, replace $\thetafn{ ( 1 -\Tr y_\mu y_\mu )} = \int_0^1 dt
\deltafn{ (t-y_\mu y_\mu)}$ and rescale $t=[u/R]^{2-\eta}$ and $y_\mu
= \tilde{y}_\mu [u/R]^{1-\eta /2}$ giving
\bea \C I^r_{D, G}(R)& <& R^{-(2-\eta)2(D-2)} R^{-(2-\eta)(D-1)(g-g'-1)} \nn\\
&&\times \sum_{r'} C_{r'} \int_0^R {du \over u} [u/R]^
{(2-\eta)[(D-1)g' + 3r'/2]} \nn\\
&& \times \int_{\C R_1(G')} \prod_{\nu=1}^D
d\tilde{y}_\nu \,  
\absval{\Pf^{r'}_{D,G'} 
(\tilde{y},u)} \deltafn{\left(1-\Tr \tilde{y}_\mu \tilde{y}_\mu \right)}.
\eea 
Since $u/R <1$, this can be reduced to
\beq\label{F17}
\C I^r_{D, G}(R)< R^{-(2-\eta)[2(D-2)+(D-1)(g-1)]}
\sum_{r'} C_{r'} \int_0^R {du \over u} u^
{(2-\eta)(D-1)g'} \C I^{r'}_{D,G'}.
\eeq 
We argue by induction, so assume that, for $G'$
\beq\label{F18}
\int_0^\infty dR R^{Dg'-1} R^{(D-2)g'} \C I^r_{D,G'}(R)
\eeq
converges for $D > 3$, and all choices of $r$. Then \ref{F17} gives
\beq
\C I^r_{D, G}(R)< C R^{-(2-\eta)[2(D-2)+(D-1)(g-1)]}
\eeq
and in particular, the induction statement is true also for $G$. It
remains to check that the induction statement is true for the smallest
possible regularly embedded subalgebra, which is $su(2)$. Since
$su(2)$ has no regularly embedded subalgebra, we can repeat the above
arguments with $\C G' =0$ and find
\beq
\C I^r_{D, SU(2)}(R)< C R^{-(2-\eta)[2(D-2)+2(D-1)]}
\eeq
and this completes the proof.

Taking now $r=0$, we have discovered that, for any compact semi-simple
group $G$, 
\beq
\C I_{D, G}(R)< C R^{-(2-\eta)[2(D-2)+(D-1)(g-1)]}\label{ssbound}
\eeq
and in particular, the partition function $\C Z_{D,G}$ converges for
$D > 3$. It is a remarkable fact that the bound \ref{ssbound} does not 
depend on the sub-algebra.  For the correlation function \ref{correl} to converge, we
require
\beq
Dg + (D-2)g +k < 2[2(D-2)+(D-1)(g-1)]
\eeq
and so the critical value is
\beq
k_c = 2(D-3)
\eeq
independently of the gauge group. 

\section{Discussion}
For the bosonic theories, we have shown that the partition function
converges when $D \geq D_c$ and calculated $D_c$ for each of the
compact simple groups. It is a simple exercise to extend the result to
the compact semi-simple groups since they are built out of the simple
groups. For example, $so(4)= su(2) \oplus su(2)$, so $\C Z_{D,SO(4)}$
converges when $D \geq D_c=5$. In addition, we have calculated the
critical degree $k_c$ for correlation functions, such that $\langle \C
C_k \rangle$ converges when $k<k_c$. Conversely, we have shown that
there always 
exists a correlation function of degree $k_c$ which diverges.

It happens that, for the simple groups, the result $D_c$ is equal
to the result one would obtain by assuming every $(x \cdot \alpha)^2$
is bigger than a constant in equation \ref{B15} (the one loop
approximation). Thus, 
in this case, there is a simple rule
\beq
D > {2(g-l) \over g-2l}
\eeq
for the partition function to converge. However, this rule fails for
the semi-simple groups (as we quickly see by considering $su(r+1)
\oplus su(2)$ for large $r$). In general, the one loop approximation
gives 
the wrong value for $k_c$ (except in the case of $su(2)$ where it is exact).

We have shown that the supersymmetric partition function
converges in $D=4,6$ and $10$ with any compact semi-simple gauge
group, and that correlation functions of degree
\beq
k<k_c =2(D-3)
\eeq
are convergent independent of the gauge group. In the case of
$SU(r+1)$, this result corresponds to a conjecture
\cite{Krauth:1999qw} based on Monte Carlo evaluation of the integrals
for small  $r$. We have not found rigorous lower bounds in the 
supersymmetric case so it remains unproven that $k_c$ is critical;
however, there can be little doubt that, for example, $\expect{\Tr X_\mu^2}$
is logarithmically divergent for $D=4$.
The situation for partition functions in $D=3$ may be more complicated;
the integrals are absolutely divergent but for some groups the Pfaffian
is an odd function of $X_\mu$ so that the integral vanishes naively.
Only if there were a supersymmetric regularization of the integrals
would it be possible that some of these theories make sense.

\acknowledgments
PA thanks M. Staudacher for his hospitality and  useful discussions,
 and acknowledges a
PPARC studentship. We are grateful to B. Durhuus, T. Jonsson, and
M. Staudacher for their helpful comments on  the first draft of this
paper.

\bibliographystyle{JHEP}
\bibliography{/home/tackley/jfw/researcher/papers/YMconvergence/YMrefs}






\end{document}